\documentclass[twoside,twocolumn,9pt]{article}
\usepackage{extsizes}
\usepackage[super,sort&compress,comma]{natbib} 
\usepackage[version=3]{mhchem}
\usepackage[left=1.5cm, right=1.5cm, top=1.785cm, bottom=2.0cm]{geometry}
\usepackage{balance}
\usepackage{mathptmx}
\usepackage{sectsty}
\usepackage{graphicx} 
\usepackage{lastpage}
\usepackage[format=plain,justification=justified,singlelinecheck=false,font={stretch=1.125,small,sf},labelfont=bf,labelsep=space]{caption}
\usepackage{float}
\usepackage{fancyhdr}
\usepackage{fnpos}
\usepackage{siunitx}
\usepackage[english]{babel}
\addto{\captionsenglish}{%
  
}
\usepackage{array}
\usepackage{droidsans}
\usepackage{charter}
\usepackage[T1]{fontenc}
\usepackage[usenames,dvipsnames]{xcolor}
\usepackage{setspace}
\usepackage[compact]{titlesec}
\usepackage{hyperref}

\definecolor{gold}{rgb}{0.85,.66,0}
\definecolor{plum}{rgb}{0.45,0,.66}
\definecolor{ngn}{rgb}{0,1,0}  
\definecolor{myred}{cmyk}{0.000000,1.000000,1.000000,0.1}
\definecolor{myblue}{cmyk}{1.000000,0.750000,0.000000,0.1}
\definecolor{mygrn}{cmyk}{0.750000,0.000000,1.000000,0.2}

\usepackage{epstopdf}

\definecolor{cream}{RGB}{222,217,201}

\begin{document}

\pagestyle{fancy}
\thispagestyle{plain}
\fancypagestyle{plain}{
\renewcommand{\headrulewidth}{0pt}
}

\makeFNbottom
\makeatletter
\renewcommand\LARGE{\@setfontsize\LARGE{15pt}{17}}
\renewcommand\Large{\@setfontsize\Large{12pt}{14}}
\renewcommand\large{\@setfontsize\large{10pt}{12}}
\renewcommand\footnotesize{\@setfontsize\footnotesize{7pt}{10}}
\makeatother

\renewcommand{\thefootnote}{\fnsymbol{footnote}}
\renewcommand\footnoterule{\vspace*{1pt}%
\color{cream}\hrule width 3.5in height 0.4pt \color{black}\vspace*{5pt}} 
\setcounter{secnumdepth}{5}

\makeatletter 
\renewcommand\@biblabel[1]{#1}            
\renewcommand\@makefntext[1]%
{\noindent\makebox[0pt][r]{\@thefnmark\,}#1}
\makeatother 
\renewcommand{\figurename}{\small{Fig.}~}
\sectionfont{\sffamily\Large}
\subsectionfont{\normalsize}
\subsubsectionfont{\bf}
\setstretch{1.125} 
\setlength{\skip\footins}{0.8cm}
\setlength{\footnotesep}{0.25cm}
\setlength{\jot}{10pt}
\titlespacing*{\section}{0pt}{4pt}{4pt}
\titlespacing*{\subsection}{0pt}{15pt}{1pt}

\fancyfoot{}
\fancyfoot[LO,RE]{\vspace{-7.1pt}\includegraphics[height=9pt]{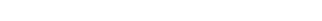}}
\fancyfoot[CO]{\vspace{-7.1pt}\hspace{13.2cm}\includegraphics{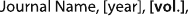}}
\fancyfoot[CE]{\vspace{-7.2pt}\hspace{-14.2cm}\includegraphics{head_foot/RF}}
\fancyfoot[RO]{\footnotesize{\sffamily{1--\pageref{LastPage} ~\textbar  \hspace{2pt}\thepage}}}
\fancyfoot[LE]{\footnotesize{\sffamily{\thepage~\textbar\hspace{3.45cm} 1--\pageref{LastPage}}}}
\fancyhead{}
\renewcommand{\headrulewidth}{0pt} 
\renewcommand{\footrulewidth}{0pt}
\setlength{\arrayrulewidth}{1pt}
\setlength{\columnsep}{6.5mm}

\makeatletter 
\newlength{\figrulesep} 
\setlength{\figrulesep}{0.5\textfloatsep} 

\newcommand{\topfigrule}{\vspace*{-1pt}%
\noindent{\color{cream}\rule[-\figrulesep]{\columnwidth}{1.5pt}} }

\newcommand{\botfigrule}{\vspace*{-2pt}%
\noindent{\color{cream}\rule[\figrulesep]{\columnwidth}{1.5pt}} }

\newcommand{\dblfigrule}{\vspace*{-1pt}%
\noindent{\color{cream}\rule[-\figrulesep]{\textwidth}{1.5pt}} }

\makeatother

\twocolumn[
  \begin{@twocolumnfalse}
{\includegraphics[height=30pt]{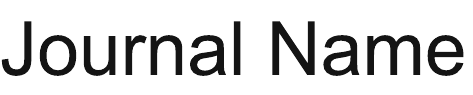}\hfill\raisebox{0pt}[0pt][0pt]{\includegraphics[height=55pt]{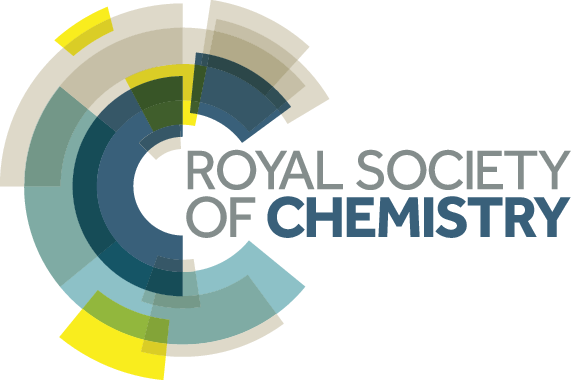}}\\[1ex]
\includegraphics[width=18.5cm]{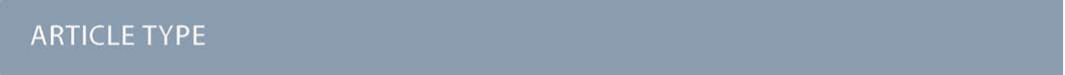}}\par
\vspace{1em}
\sffamily
\begin{tabular}{m{4.5cm} p{13.5cm} }

\includegraphics{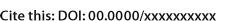} & \noindent\LARGE{\textbf{Leveraging mid-infrared spectroscopic imaging and deep learning for tissue subtype classification in ovarian cancer$^\dag$}} \\
\vspace{0.3cm} & \vspace{0.3cm} \\

\author{Chalapathi Charan Gajjela}, 
\author{Matthew Brun},
\author{Rupali Mankar},
\author{Noah Kennedy},
\author{Sara Corvigno},
\author{Yanping Zhong},
\author{Jinsong Liu},
\author{Anil K. Sood},
\author{David Mayerich},
\author{Sebastian Berisha},
\author{Rohith Reddy}*

 & \noindent\large{Chalapathi Charan Gajjela,\textit{$^{a}$} Matthew Brun,\textit{$^{b}$} Rupali Mankar,\textit{$^{a}$} Sara Corvigno,\textit{$^{d}$} Noah Kennedy,\textit{$^{c}$}  Yanping Zhong,\textit{$^{d}$} Jinsong Liu,\textit{$^{d}$} Anil K. Sood,\textit{$^{d}$} David Mayerich,\textit{$^{a}$} Sebastian Berisha,\textit{$^{c}$} and Rohith Reddy$^{\ast}$\textit{$^{a}$}} \\

\includegraphics{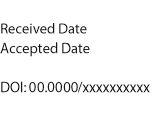} & \noindent\normalsize{Mid-infrared spectroscopic imaging (MIRSI) is an emerging class of label-free techniques being leveraged for digital histopathology. Modern histopathologic identification of ovarian cancer involves tissue staining followed by morphological pattern recognition. This process is time-consuming, subjective, and requires extensive expertise. This paper presents the first label-free, quantitative, and automated histological recognition of ovarian tissue subtypes using a new MIRSI technique. This technique, called optical photothermal infrared (O-PTIR) imaging, provides a 10$\times$ enhancement in spatial resolution relative to prior instruments. It enables sub-cellular spectroscopic investigation of tissue at biochemically important fingerprint wavelengths. We demonstrate that enhanced resolution of sub-cellular features, combined with spectroscopic information, enables reliable classification of ovarian cell subtypes achieving a classification accuracy of $0.98$. Moreover, we present statistically robust validation from 74 patient samples with over 60 million data points. We show that sub-cellular resolution from five wavenumbers is sufficient to outperform state-of-the-art diffraction-limited techniques from up to 235 wavenumbers. We also propose two quantitative biomarkers based on the relative quantities of epithelium and stroma that exhibits efficacy in early cancer diagnosis. This paper demonstrates that combining deep learning with intrinsic biochemical MIRSI measurements enables quantitative evaluation of cancerous tissue, improving the rigor and reproducibility of histopathology.} \\

\end{tabular}
 \end{@twocolumnfalse} \vspace{0.6cm}

  ]

\renewcommand*\rmdefault{bch}\normalfont\upshape
\rmfamily
\section*{}
\vspace{-1cm}


\footnotetext{\textit{$^{a}$~Address, 4226 Martin Luther King Boulevard,  N308 Engineering Building 1, Houston TX, 77584, USA; E-mail: rkreddy@uh.edu}}




\section{Introduction}
Epithelial ovarian cancer is the leading cause of death among gynecological malignancies in the United States. Serous ovarian cancer, its most common subtype, is often diagnosed at a late stage (III or IV), where 5-year survival is 51\% and 29\%, respectively.\cite{torre2018ovarian}  Standard treatments involve surgery and at least six courses of chemotherapy.\cite{lheureux2019epithelial} Several novel compounds have been studied and approved over the past 20 years; however, none substantially modify overall survival.~\cite{lee2019new} The most decisive prognostic factor remains the complete eradication of neoplastic tissue through radical surgery.~\cite{du2009role, kehoe2015primary, vergote2010neoadjuvant} Outcomes are affected by  (1) late diagnosis resulting in unresectable disease and (2) unclear identification of neoplastic margins. Therefore, objective and early identification of neoplastic tissue are essential for optimal surgical attempts.

Recent advances reveal the complex organization of the ovarian tumor microenvironment, highlighting inter-cellular pathways~\cite{luo2016tumor} as potential treatment targets. New methods quantifying biomolecular characteristics reveal detailed structural and molecular changes that may reveal novel therapeutic targets.  
The current standard for ovarian cancer diagnosis uses contrast-inducing stains on biopsy sections followed by microscopic examination by a pathologist. Hematoxylin and eosin (H\&E) stain is widely used to identify cellular and extracellular components. {\it Epithelial} carcinoma is the most common histologic type, accounting for about 90 percent of cancers of the ovary, fallopian tube, and peritoneum.~\cite{banks_epidemiology_2001, heintz_carcinoma_2001} In high-grade serous carcinoma (HGSC), a pathologist identifies various architectural patterns, including complex papillary, glandular, microcystic, and solid patterns. HGSC infiltrates, destroys, and/or replaces the normal {\it stroma}. Therefore, histological identification of cellular subtypes is an important step~\cite{malpica2008grading} in ovarian cancer diagnosis and prognosis.   

Inter-pathologist variability is a significant challenge, \cite{hernandez1984interobserver} and grading schemes have been proposed to reduce this variability.~\cite{malpica2004grading, taylor1999analysis, zeppernick2014new} However, these methods have only been successful in resource-rich hospitals with comprehensive training.~\cite{malpica2007interobserver} Automated and semi-automated techniques to reduce inter-pathologist variability, especially in lower-resource settings, are critical for equitable care.~\cite{bera2019artificial, wu2018automatic, komura2018machine} Automated tissue subtyping is an essential step in this effort.     

Automated tissue classification into the epithelium and stroma subtypes is challenging, and several techniques have been proposed. Most use H\&E~\cite{du2018classification, chen2020utilizing, xu2017deep} and immunohistochemical staining~\cite{fiore2012utility} combined with machine learning (ML). Staining quality and variability can confound ML and lead to inconsistent results.~\cite{wu2018automatic} Our goal is to perform label-free recognition of tissue subtypes without using chemical contrast agents. Moreover, we obtain intrinsic quantitative and repeatable biochemical measurements that are independent of operator tissue processing.   

Spectroscopic techniques are used widely to identify molecules in chemical and biochemical analysis \cite{ur2012vibrational} with high sensitivity and specificity. Vibrational spectroscopy is used routinely to identify organic biomolecules by matching measurements to large commercial spectral libraries containing over 260,000 spectra.\cite{d2009extracting}  
%
Prior work on label-free ovarian tissue analysis has utilized spectroscopy. Raman spectroscopy,\cite{morais2019three, paraskevaidi2018raman, maheedhar2008diagnosis} conventional Fourier Transform Infrared (FTIR) spectroscopy,\cite{krishna2007ftir, flower2011synchrotron, gajjar2013fourier} and attenuated total reflection (ATR) FTIR spectroscopy~\cite{theophilou2016atr, lima2015segregation} have been applied to detect and diagnose ovarian cancer. However, prior work lacked spatial specificity and required long data collection times. MALDI imaging~\cite{klein2019maldi} has also been used to analyze ovarian histotypes; however, this technique destroys the sample. 
Second-harmonic generation (SHG) can identify collagen in the stroma,~\cite{zeitoune2017epithelial, pouli2019two, tilbury2015applications} and multi-photon microscopy~\cite{huttunen2018automated} has been used on murine tissue.
Raman imaging has been used for ovarian cancer diagnosis and tissue analysis, often with added nanoparticles~\cite{jokerst2012gold,oseledchyk2017folate} to obtain more robust signals. Taken together, none of these techniques provide classified images in an automated, label-free, quantitative, and non-destructive manner.

Mid-infrared spectroscopic imaging (MIRSI) can extract spectral and spatial information without using contrast agents by utilizing intrinsic biochemical properties of tissue. This technology is non-destructive and therefore compatible with other technologies.~\cite{petibois2006chemical} Fourier transform infrared (FTIR) spectroscopic imaging, the best known MIRSI technology, can classify cell subtypes in a variety of diseases, including breast,~\cite{benard2014infrared} lung,~\cite{grosserueschkamp2015marker} prostate,~\cite{baker2009investigating} and colon~\cite{krafft2008raman} cancers. We hypothesize that it is also helpful for ovarian cancer tissue subtyping.  HGSC is the marked cytologic atypia with prominent mitotic activity in ovarian tissue. The atypical nuclei are hyperchromatic with an over threefold variation in nuclear size. Phosphate spectroscopic peaks (1080, 1201, 1236, 1262 cm$^{-1}$) corresponding to nucleic acids strongly correlate to mitotic activity, and spectroscopic imaging has shown increased phosphate signals in a variety of cancers.\cite{baker2014using, pounder_development_2016, pahlow_application_2018}. 

\begin{figure}[h]
    \centering
		\includegraphics[width=\linewidth]{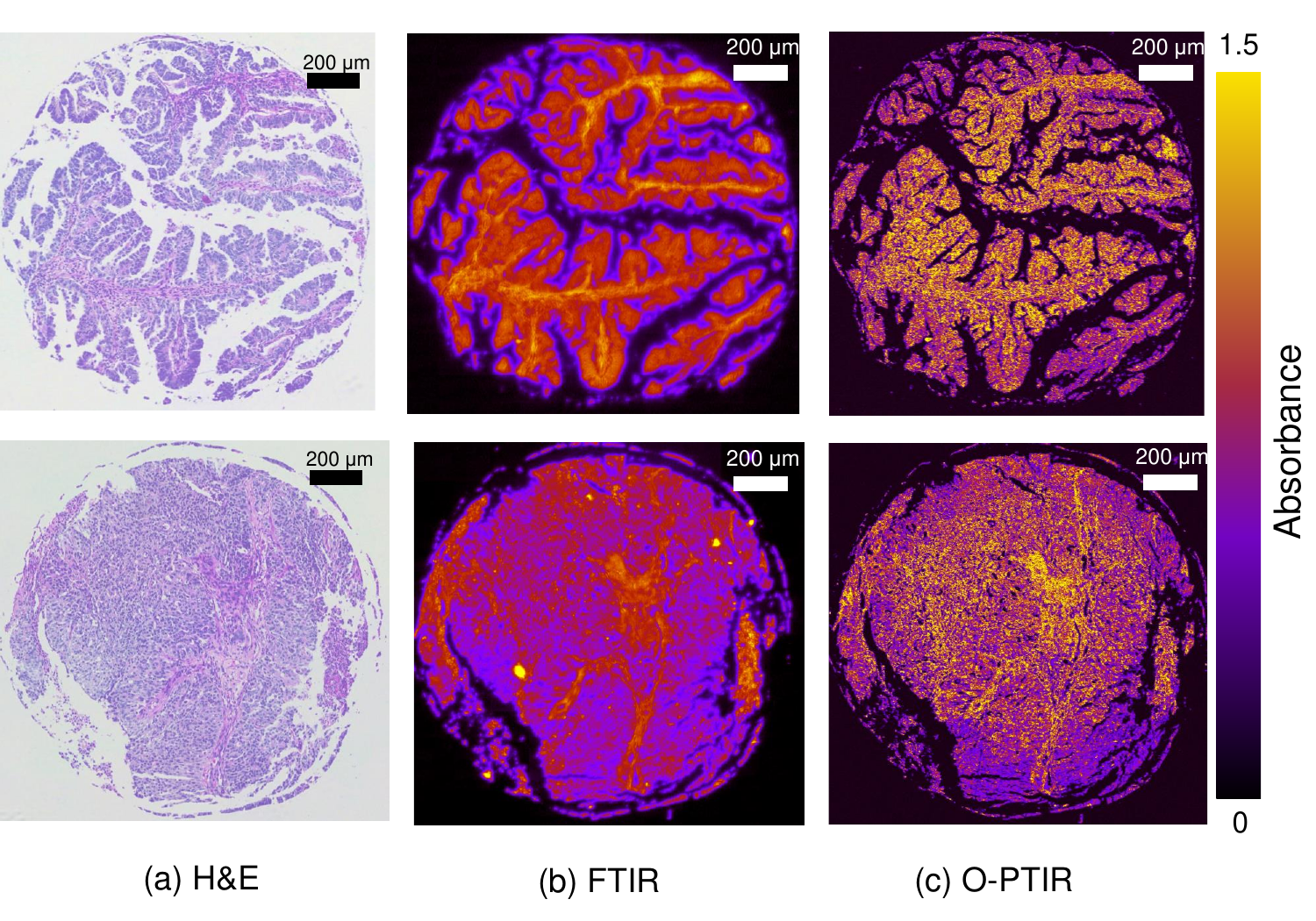}
	\caption{Two tissue cores from ovarian cancer patients are stained with H\&E, and their microscopy images are presented in (a). Corresponding tissue images at \SI{1664}{\per\centi\meter} from FT-IR are presented in (b), and data from O-PTIR are in (c). The Figure demonstrates a good correspondence between H\&E  and spectroscopic imaging data. Moreover, the O-PTIR images have a higher resolution, and finer tissue details are visible in (c) relative to images in (b). }
	\label{fig:stains-optir}
\end{figure}

FTIR imaging is limited by the diffraction of mid-infrared light (\SI{2.5}{\micro\meter} - \SI{11}{\micro\meter}). Since typical cells are of $~5 \mu$m in size, FTIR imaging cannot provide sub-cellular information potentially important for analysis. Optical-photothermal infrared (O-PTIR) imaging overcomes this resolution limitation. This technique combines a visible laser beam and a mid-infrared beam in a pump-probe architecture and estimates the sample's absorbance by measuring the change in intensity of the visible laser caused by the photothermal effect. Therefore the image resolution is determined by the wavelength of visible light (\SI{0.5}{\micro\meter}), which is much shorter than the wavelength of IR light incident on the sample, allowing $5$x to $22$x improvement in spatial resolution.~\cite{zhang2016depth, mankar2022polarization} The improved spatial resolution is comparable to an optical microscope image, as demonstrated in Figure \ref{fig:stains-optir}. This technology has been used previously for studying the chemistry of  inorganic 2D perovskite and allowed us to understand the edge emission phenomenon in inorganic 2D perovskite.~\cite{qin2020spontaneous} O-PTIR has also been successfully used in chemical imaging of live human ovarian cancer cells,~\cite{bai2019ultrafast} where high resolution is needed to analyze sub-cellular structures in a small area. However, our work is the first large-scale (74 cancer patients) study of clinical ovarian tissue biopsy samples, each with large sample areas (1 mm diameter each).  

Machine learning algorithms, including random forest (RF) and Bayesian classifiers, use individual spectra to classify tissue from MIRSI data.  These approaches do not leverage spatial information, although MIRSI provides both spatial and spectroscopic information. Convolutional neural networks (CNNs) are deep learning architectures that  learn local spatial features. CNNs have been successful in hyperspectral \cite{hu2015deep} and mid-infrared spectroscopic classification of breast cancer tissues.~\cite{berisha2019deep} Traditional CNNs consist of alternating convolution and pooling layers, followed by a fully-connected classifier. In this paper, we use CNNs to determine the impact of improved O-PTIR resolution on classification.  Previous work in ovarian cancer spectroscopy relied on point spectra to identify  malignant tissue or grade tumors.~\cite{theophilou2016atr, maheedhar2008diagnosis}  We report the first application of the O-PTIR tissue classification and the first reported application of MIRSI and deep learning to ovarian histology. 

\subsection{O-PTIR imaging}
\begin{figure}[hbt]
    \centering
  \includegraphics[width=0.8\linewidth]{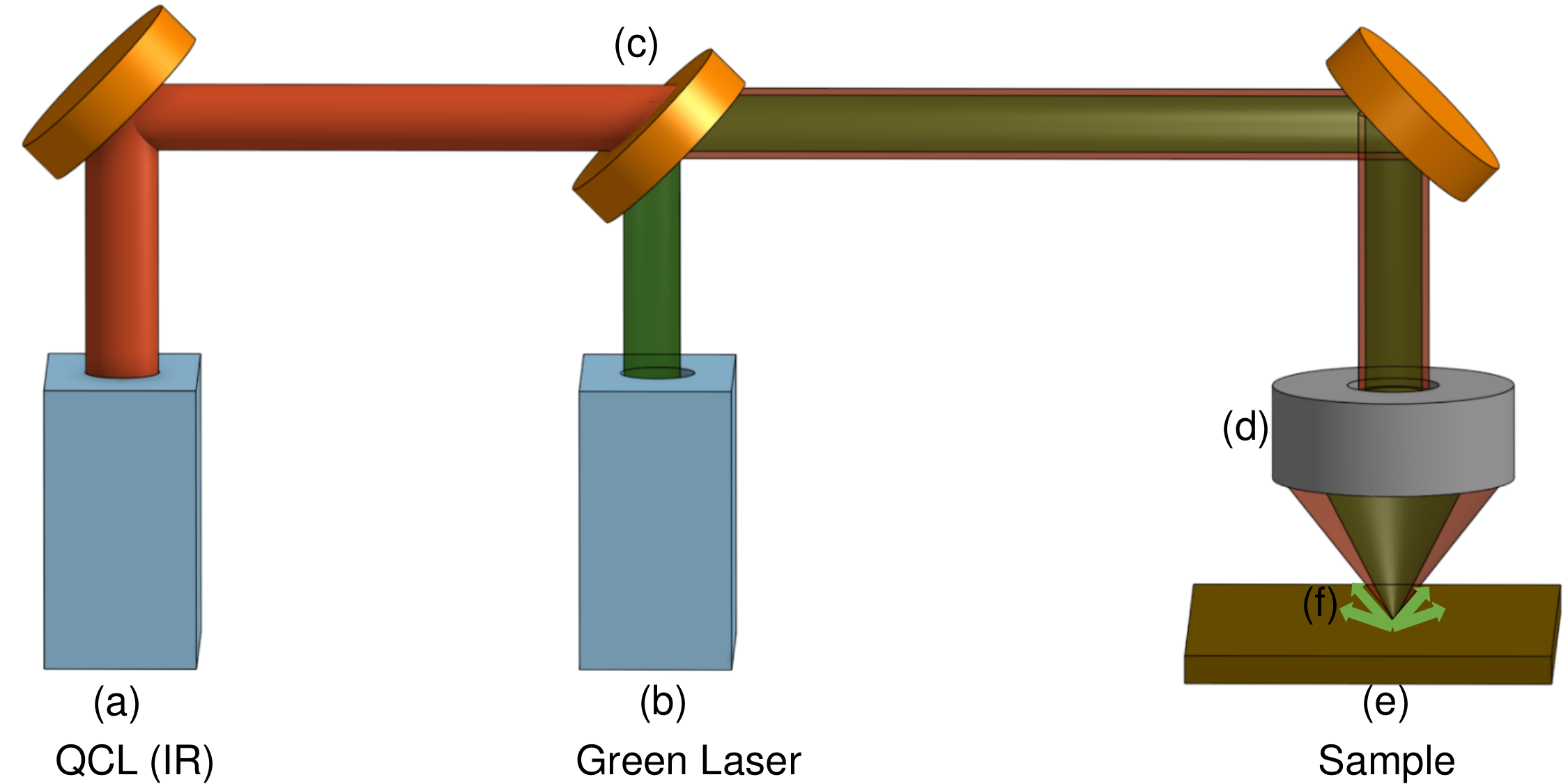}
    \caption{Schematic of the optical path of the IR and green (532 nm) lasers in our O-PTIR instrument.  A pulsed Quantum Cascade Laser (QCL) shown in (a) is the source of mid-IR light which acts as a pump causing photothermal expansion at the sample. A Continuous Wave (CW) green laser shown in (b) is incident collinearly on the sample and acts as a probe beam. A dichroic mirror (c) combines the green and QCL light and focuses them on the sample (e) using a reflective Cassegrain objective (d). The modulation in the intensity of the green light (f) scattered back from the sample enables the measurement of its IR absorbance. }
    \label{fig:schematic}
\end{figure}

Photothermal microscopy obtains a measurement of sample absorbance by estimating the thermal expansion caused by the absorption of infrared light using a co-localized visible laser beam. The visible and IR laser are incident on the sample collinearly, as shown in Figure\ref{fig:schematic}. The thermal expansion caused in the sample due to IR absorbance causes variation in the refractive index due to the photothermal effect. This change in refractive index is detected by measuring the change in intensity of the back-reflected green laser (visible laser) using a point detector. 
\subsubsection{Improved Spatial Resolution}
\begin{figure}[hbt]
    \centering
  \includegraphics[width=\linewidth]{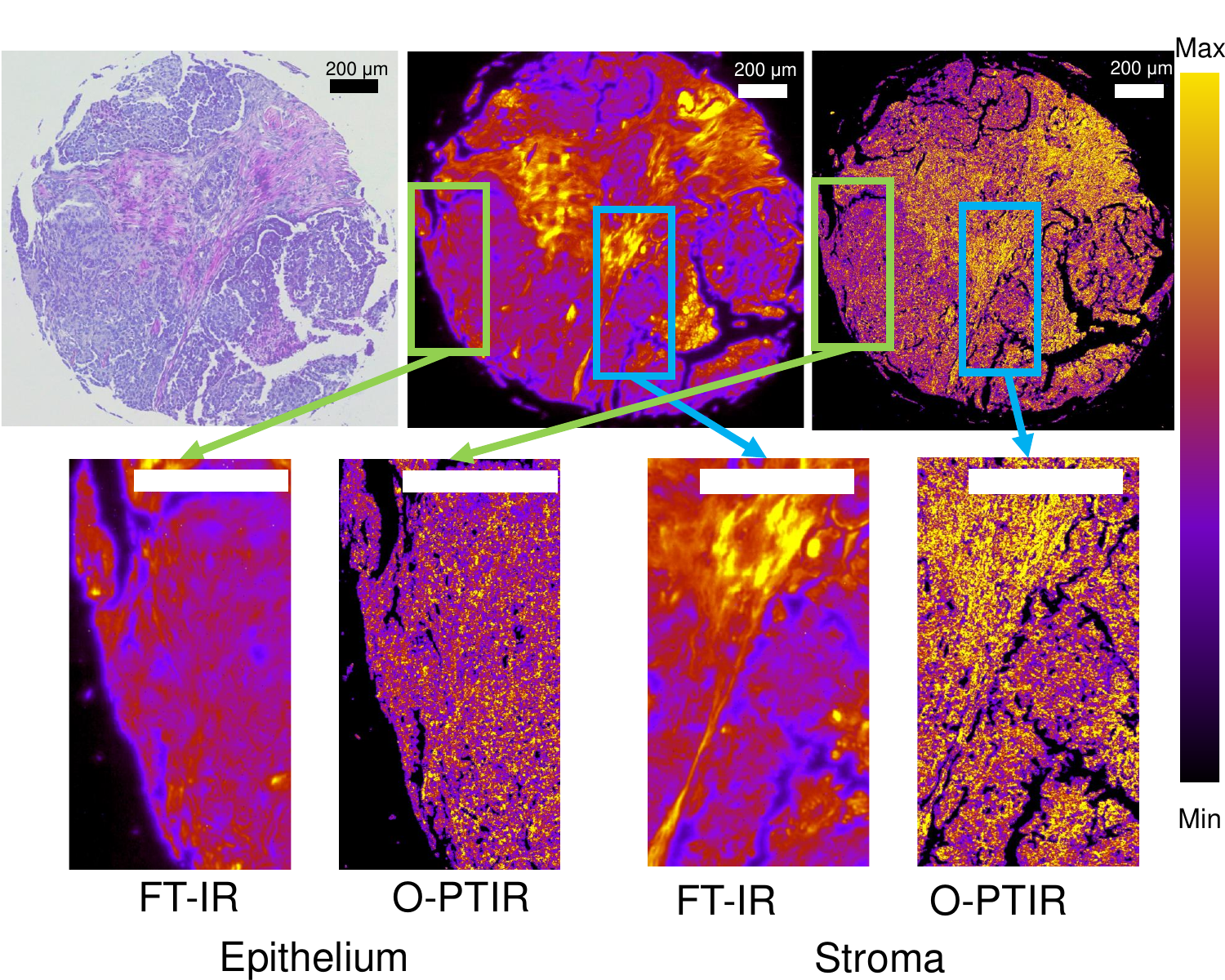}
    \caption{Spatial differences between different cell types in FT-IR image on the left, H \& E image in the center and O-PTIR image on the right. Cropped regions around pixels from the same core in FT-IR and O-PTIR images collected at 1650 $cm^{-1}$. }
    \label{fig:classes}
\end{figure}

FTIR has been the standard spectroscopic imaging technique for characterizing a material's chemistry.  While FTIR provides spectral data across all mid-infrared wavenumbers, its spatial resolution is diffraction-limited by the long wavelengths of light resulting in modest image quality. O-PTIR overcomes this limitation and provides higher spatial resolution (\SI{0.5}{\micro\meter}) images, with data quality comparable to H\&E stained microscopy images. Figure \ref{fig:classes} compares the image quality of O-PTIR, FTIR, and H\&E on the same cancer tissue. The improved spatial resolution of O-PTIR relative to FTIR is evident. Furthermore, O-PTIR data quality is comparable to microscopy data after H\&E staining on an adjacent section. Finer spatial details in the epithelium and stromal tissue regions are also observed in the O-PTIR data presented in Figure \ref{fig:classes}. \par

\section{Materials and Methods}
An ovarian biopsy tissue microarray (TMA) was obtained from Biomax US, Rockville, MD (TMA ID: BC11115c) and imaged using O-PTIR. The TMA consists of 100 cores from separate patients with cases of normal, hyperplastic, dysplastic, and malignant tumors. The patient cohort was composed of women aged 29 to 69; ovarian tumor stages varied between stage I to stage IIIC; histological subtypes include clear cell carcinoma, high-grade serous carcinoma, and Mucinous adenocarcinoma. The paraffin-embedded samples were deparaffinized before undergoing O-PTIR imaging. The corresponding adjacent histological section was stained with H\&E and examined by an expert pathologist. Cell subtypes were identified across disease stages. We trained a random forest (RF) classifier, support vector machine (SVM), k-nearest neighbor (KNN), and a CNN model on randomly selected pixels from half of the TMA cores, using the other half as validation.

\subsection{Data acquisition}
FTIR images were acquired using an Agilent Stingray imaging system equipped with a $680$-IR spectrometer connected to a $620$-IR imaging microscope with a numerical aperture of $0.62$. Each core was imaged with $16$ co-additions in transmission mode at a spectral resolution of \SI{8} {\per\centi\meter} truncated from \SI{902} {\per\centi\meter} to \SI{3892} {\per\centi\meter}, and a pixel size of \SI{1.1}{\micro\meter}.We collected the background scan at $128$ co-additions and ratioed to the single beam data to remove spectral contributions from the substrate, atmosphere, and globar source.  \par
\begin{figure}[hbt]
  \includegraphics[width=\linewidth]{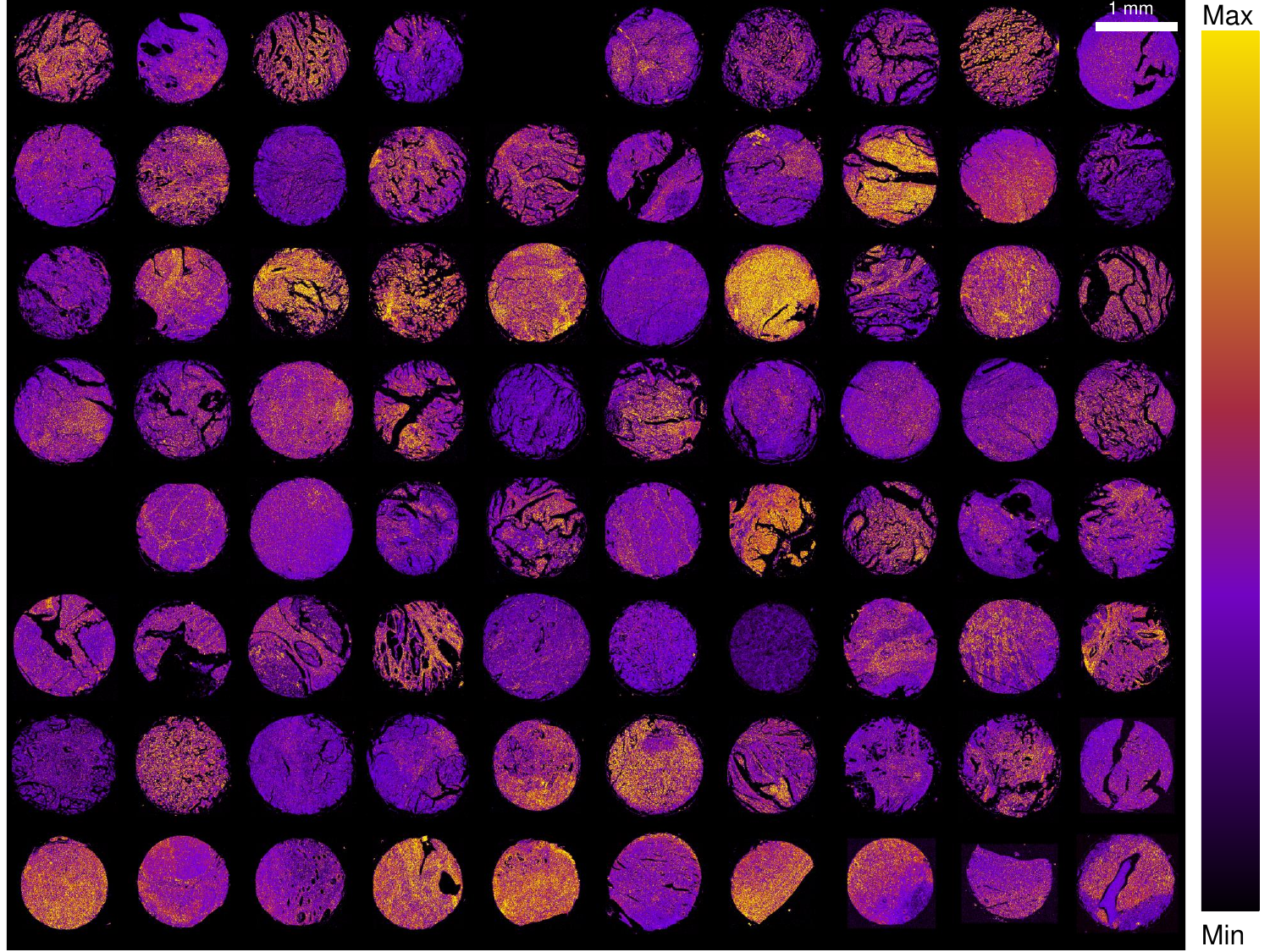}
    \caption{O-PTIR microarray (8$\times$10) shown at band \SI{1664}{\per\centi\meter}. Data from 78 ovarian cancer patients is shown. The biochemical variations in tissue are evident from the differences in color in the Figure.  Machine learning algorithms combine biochemical data from multiple bands enabling tissue subtype identification and disease diagnosis.}
    \label{fig:ir-tma}
\end{figure}

The O-PTIR dataset was acquired using a Photothermal mIRage microscope with a silicon photodiode, a pixel size of \SI{0.5}{\micro\meter}$\times$\SI{0.5}{\micro\meter} and a $0.65$ numerical aperture. A Quantum Cascade Laser (QCL) source sweeps through the range of \SI{902} {\per\centi\meter} to \SI{1898} {\per\centi\meter}. Each core was imaged at five selected wavenumbers (\SI{1162}{\per\centi\meter}, \SI{1234}{\per\centi\meter}, \SI{1396}{\per\centi\meter}, \SI{1540}{\per\centi\meter}, and \SI{1661}{\per\centi\meter} ). An image of the entire TMA acquired at the Amide I band is shown in Figure \ref{fig:ir-tma}. 
Background spectra are collected with 10 co-additions and used to normalize the raw data to calculate the IR absorbance at each band. Band images are then normalized with Amide I (\SI{1664}{\per\centi\meter}) to bring the data range between $0$ and $1$. Note that some tissue biopsies are missing because the collected data was corrupted. 
We used an Aperio Scanscope system to acquire the Light microscope images of the whole slide's H\&E chemically stained sections.
\subsection{Feature Selection}
\begin{figure}[hbt]
  \includegraphics[width=\linewidth]{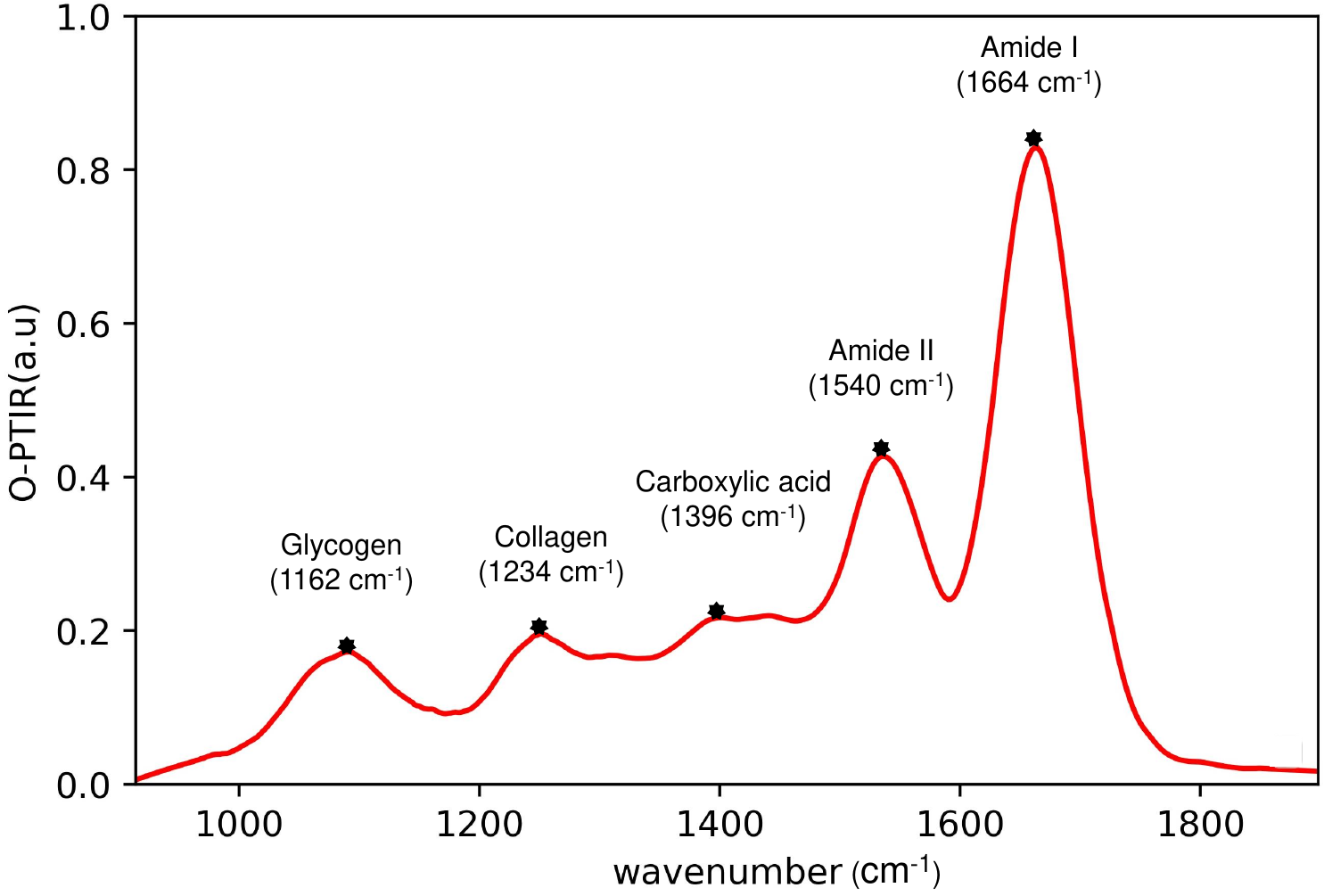}
    \caption{The Figure shows the absorption spectrum of ovarian tissue collected using O-PTIR. The spectrum shows the IR absorption values (Y-axis) for the wavenumbers (X-axis) in the fingerprint region. The marked peaks on the spectrum correspond to biochemically relevant functional groups~\cite{zohdi2015importance} of glycogen at \SI{1162}{\per\centi\meter}, amide III at \SI{1234}{\per\centi\meter}, nucleic acids and lipids at \SI{1396}{\per\centi\meter}, amide II at \SI{1540}{\per\centi\meter}, and amide I at \SI{1661}{\per\centi\meter}.}
    \label{fig:spectrum}
\end{figure}

Since the O-PTIR signal is detected using a point detector, the time taken to collect an image of a single core varies between 90 to 100 minutes per wavenumber.\cite{lotfollahi2020adaptive} Collecting a hyperspectral data cube of a core at all wavenumbers (\SI{900}{\per\centi\meter} to \SI{1900}{\per\centi\meter} at \SI{2}{\per\centi\meter} spacing) would take approximately 35 day. 
We, therefore, collect fewer bands, focusing on acquiring important biochemical information. These wavenumbers are determined by analyzing FTIR spectra of ovarian tissue to determine absorbance peaks corresponding to functional groups relevant to ovarian tissue analysis based on prior work. We acquired O-PTIR data at wavenumbers \SI{1162}{\per\centi\meter}, \SI{1234}{\per\centi\meter}, \SI{1396}{\per\centi\meter}, \SI{1540}{\per\centi\meter}, and \SI{1661}{\per\centi\meter} (Figure \ref{fig:spectrum}), which correspond to glycogen, amide III, nucleic acids and lipids, amide II, and amide I, respectively.~\cite{zohdi2015importance, baker2014using}

\subsection{Data annotation}
Two pathologists labeled areas in tissue cores as \textit{stroma} or \textit{epithelium} using H\&E stained microscopy data. H\&E with IR images were aligned manually, and the labels were transferred to O-PTIR images to create annotated data for machine learning. The TMA was divided into two halves. The right half was used for training and the left for testing, with an equal number of cores in each cohort. 

\subsection{Classification Models and Hyperparameters}

The SVM classifier was trained on $10,000$ randomly selected pixels per class from the training dataset (Table \ref{tbl:pixelct}).  An equal number of data points are drawn from each class to balance the training data and optimize classifier performance. The RF classifier was trained with $100$ trees using 10,000 samples per class. Classifier inputs consisted of five-element vectors containing the IR absorption values at each pixel and the corresponding pixel label. 

\begin{figure}[hbt]
  \includegraphics[width=\linewidth]{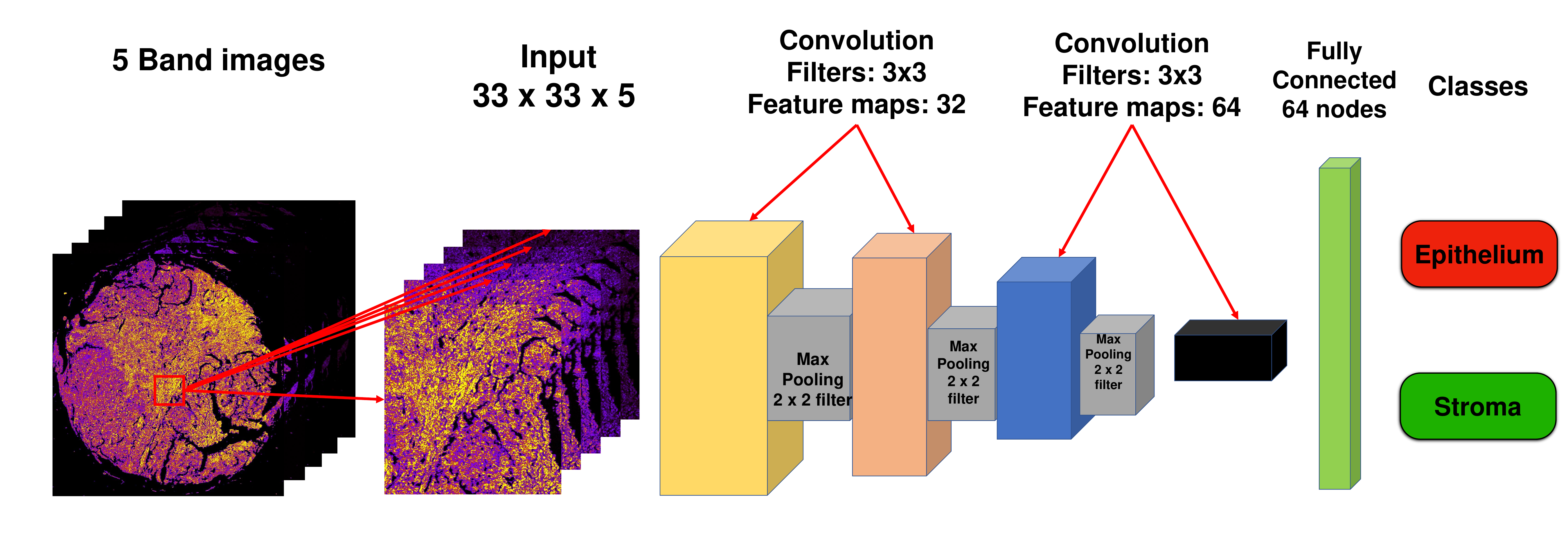}
    \caption{Schematic of the CNN architecture used for classification of O-PTIR data. A spatial region of size $33\times33$ is cropped around each pixel. Data cubes of size $33\times 33\times\times 5$ are fed into the first convolution layer. Each input is convolved with filters of size $3\times3$ outputting 32 feature maps. The following layer is a max pooling layer, which reduces the spatial dimensions by half. Feature extraction continues with three more convolution layers consisting of one $32$ and two $64$ feature maps consecutively. The extracted features are then flattened and fed to a fully connected layer with $128$ units. The last layer, softmax, consisting of $2$ units (number of classes) outputs a vector of class probabilities. In the end, the maximum probability is used to map each input pixel to its corresponding class labels.}
    \label{fig:network}
\end{figure}

\begin{table}[hbt]
\centering
 \caption{The total number of O-PTIR pixels in the training and testing datasets separated by class is presented. The TMA is split in half to create the training and testing cohorts. A small, random data subset is chosen during the first training step, and the classifier is optimized. Equal numbers of pixels are selected from each class to prevent class bias in training. $10,000$ O-PTIR pixels per class are used in the SVM and RF classifiers and $400,000$ pixels per class for CNNs.}
    \label{tbl:pixelct}
  \begin{tabular}[htbp]{@{}lll@{}}
    \hline
    Class & Training & Testing\\
    \hline
    Epithelium & $22,766,257$ & $19,249,625$\\
    Stroma & $11,001,575$ & $8,719,719$\\
    \hline\hline
    Total & $33,767,832$ & $27,969,344$\\
    \hline
  \end{tabular}
\end{table}
The CNN model uses the same general structure as our previous breast classifier~\cite{berisha2019deep}. We optimized the network for O-PTIR data classification using the following parameters. Inputs are cropped into $32\times32$ regions around the center pixel to leverage the local spatial information. The network consisted of: (1) a convolution layer with $32$-feature maps, (2) a $2\times2$ max-pooling layer, (3) a convolution layer with $32$-feature maps, (4) a convolution layer with $64$-feature maps, (5) and finally a fully connected layer with $64$ nodes and a softmax output for class probabilities. The O-PTIR network contains an additional $32$-feature convolution layer and $2\times2$ max pooling layer before the $64$-feature convolution layers to adjust to the larger input size, ensuring that the feature size in the final layers is equivalent in both models (Figure \ref{fig:network}). The same architecture is optimized for the FTIR data to make an apt comparison of the classification results.

The following hyperparameters are used for our CNN models:
\begin{enumerate}
    \item \textbf{Optimization:} The Adam optimizer was used.\cite{kingma2019method}
    \item \textbf{Dropout:} The networks had a dropout layer before the fully connected layer with a keep probability of $0.5$. This layer aids regularization and prevents overfitting by randomly disabling nodes in the first and last layers in training. The dropout layer was included before the softmax layer and before the first $64$-feature convolution layer.
    \item \textbf{Non-linearity:} A rectified linear unit (ReLU) activation function is used for each layer.
    \item \textbf{Weight initialization:} The initial weights of keras layers are initialized using the \textit{randomnormal} class from the built-in initializer, which generates tensors with a normal distribution with mean at 0.0 and standard deviation of 0.05.
    \item \textbf{Batch Size:} The networks are trained on batches of $128$ images each of size $33\times 33\times 5$.
    \item \textbf{Epochs:} The networks were trained for $8$ epochs, with data randomly shuffled between epochs. 
\end{enumerate}

\subsection{Implementation}
All data pre-processing was performed using our open-source SIproc, software~\cite{berisha2017siproc} implemented in C++ and CUDA. Training and testing were performed in Python using open-source software packages. The CNNs were implemented in Python with the Keras library,~\cite{chollet2015keras} built on TensorFlow.~\cite{tensorflow2015whitepaper}  RF and SVM classifiers and accuracy scores were computed using the Scikit-learn library.~\cite{scikit-learn} The CNN classifier's performance was calculated by testing the classifiers on ten different sets of randomly selected training pixels and averaging the overall accuracy run on an NVIDIA Tesla K40m GPU.

\subsection{Cancer detection metrics}
We define a pair of metrics that utilize tissue subtype classification results to aid early ovarian cancer detection. The "stromal ratio" (SR) is calculated by dividing the number of pixels classified as stroma by the total number of pixels in each core. Similarly, the "epithelial ratio" (ER) is the number of epithelial pixels divided by the total number of pixels in the core. 

\section{Results}
We evaluated classifier performance using the overall accuracy (OA) and receiver operating characteristic (ROC) curves. OA is beneficial for binary and multi-class classification, representing the percentage of pixels mapped correctly to the appropriate class. The ROC curves delineate the correlation between specificity and sensitivity to ascertain acceptable false positive and true positive indicators. 

We performed tissue segmentation using multiple machine learning algorithms, including those based on spectra alone, such as RFs and SVMs, and those that utilize both spatial and spectral features, such as CNNs. The class-wise and overall accuracy obtained after the classification of the testing dataset are summarized in Table ~\ref{tbl:rf-cnn-accuracy}.  CNNs outperform RF and SVM classifiers in both class-wise and overall accuracy (by 40-50\%). The low OA scores for RF ($53.21\%$) and SVM ($45.57\%$) can be attributed to using spectral information from only 5 wavenumbers instead of all the wavenumbers in the 900-1900 $cm^{-1}$ range. Meanwhile, the high overall accuracy achieved by CNN ($94.61\%$) is due to the utilization of both spectral and spatial features.

\begin{table}[hbt!]
\centering
 \caption{Accuracy scores for epithelium and stroma classification using (a) SVM, (b) RF, and (c) CNNs averaged across 80 repetitions are presented below. CNNs utilize a combination of spatial and spectral features and outperform SVMs and RFs that employ spectral features alone.}
    \label{tbl:rf-cnn-accuracy}
  \begin{tabular}[htbp]{@{}llll@{}}
    \hline
    Class & SVM & RF & CNN \\
    \hline
    Epithelium & $80.31 \pm 0.18$ & $60.18 \pm 0.29$ & $95.33 \pm 1.52$\\
    Stroma & $29.84 \pm 0.26$ & $44.27 \pm 0.21$ & $93.00 \pm 1.97$\\
    \hline\hline
    Total & $45.57 \pm 0.3$ & $53.21 \pm 0.05$ & $94.61 \pm 0.82$\\
    \hline
  \end{tabular}
\end{table}

\begin{figure}[hbt]
    \centering
  \includegraphics[width=\linewidth]{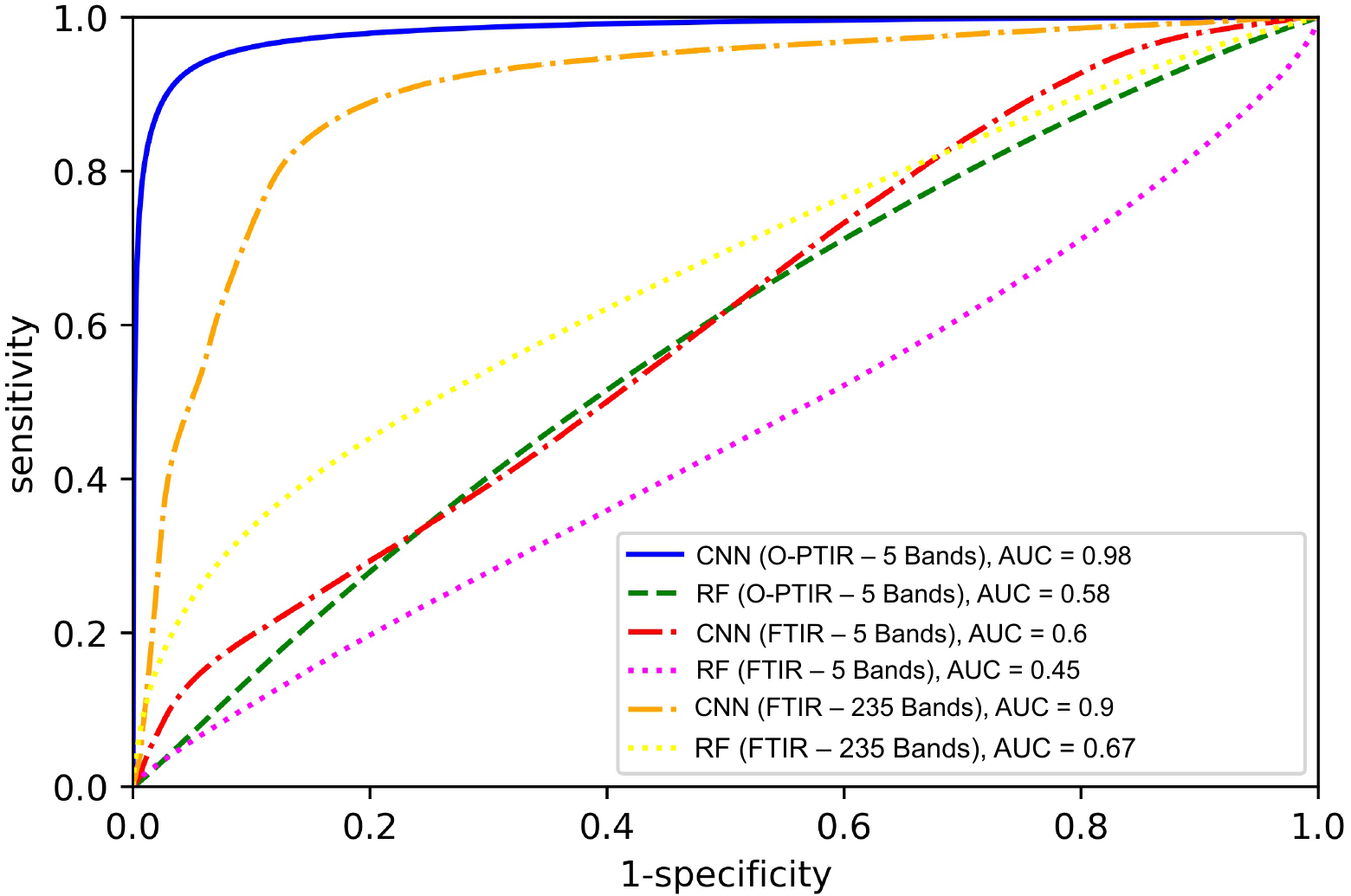}
    \caption{ROC curves and associated AUC values for binary classification of tissue type, separated by classifier type and datasets used.  Due to the use of two-class models, each tissue class curve is a reflection of the curve from the other class, and thus the AUC values are equal across tissue class.  CNN classifiers exhibit superior results to the RF classifiers, indicating that spatial information is essential in distinguishing tissue types.}
    \label{fig:roc}
\end{figure}
Results that characterize the performance of all classifiers using the Area Under the Curve (AUC) in a Receiver Operating Characteristic (ROC) plot are presented in Figure \ref{fig:roc}. Our CNN classifier on O-PTIR data from 5 bands outperforms all others with an AUC of $0.98$. An RF classifier on the same O-PTIR data shows an AUC of 0.58. Since CNNs use spatial features and RFs don't, these results highlight the importance of combining spatial and spectroscopic information for improved tissue classification. RF classification on FTIR data with 5 bands provides a poor AUC of 0.45. This AUC increases to 0.67 by incorporating the 235 fingerprint bands. Classification of FTIR data from 5 bands using a CNN yields an AUC of 0.6, which increases to 0.9 when we include the 235 fingerprint bands. Comparing CNN performance for FTIR with fingerprint spectrum (AUC=0.9) and O-PTIR with 5 bands (AUC=0.98) implies that spatial details obtained from O-PTIR  compensate for the loss in spectroscopic information due to the reduction in the number of bands. 

\begin{figure}[hbt]
    \centering
  \includegraphics[width=\linewidth]{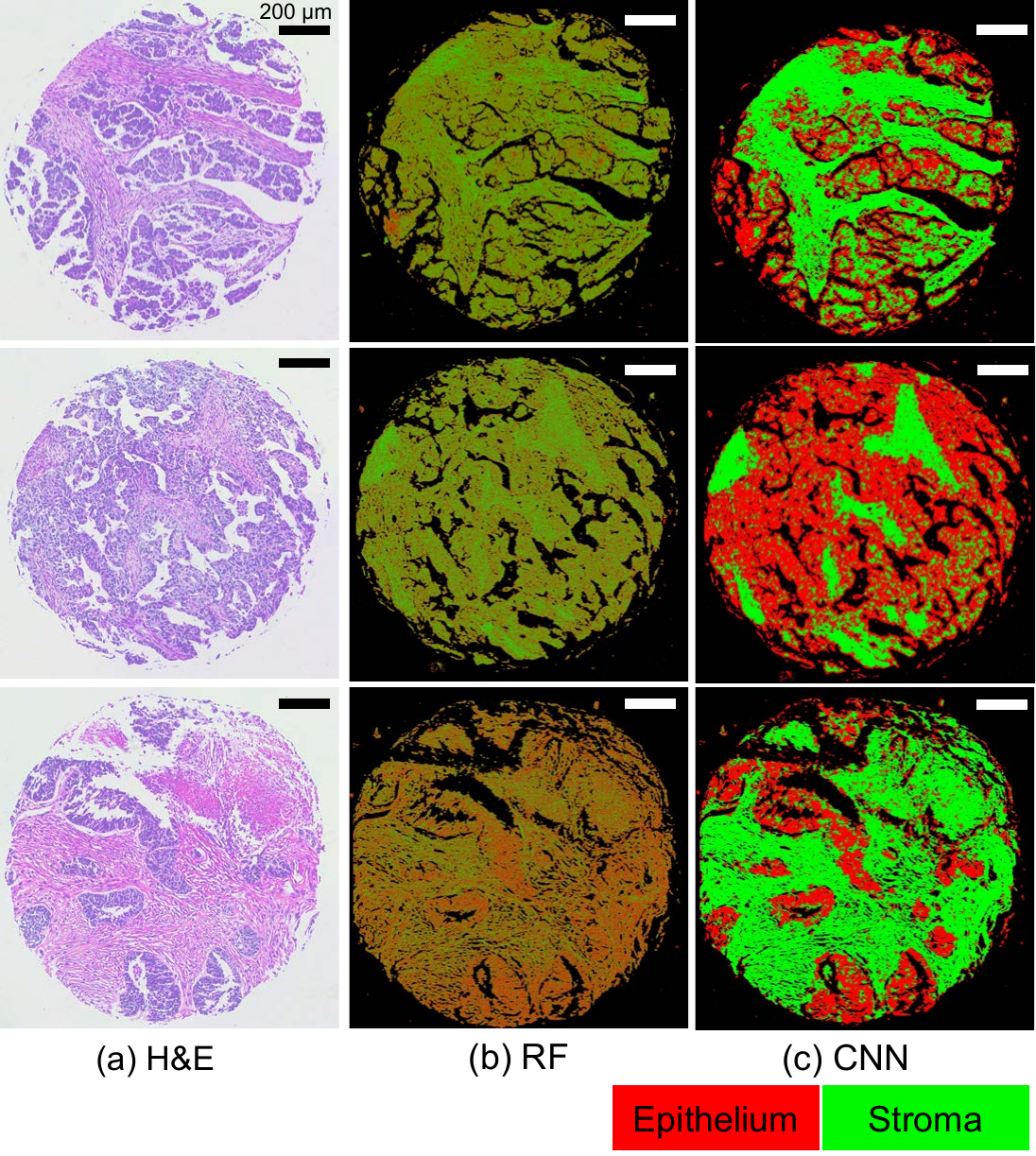}
    \caption{(a) H\&E images of tissue cores are compared to O-PTIR classification results from (b) RFs and (c) CNNs. There is a good correspondence between O-PTIR class images in (c) and the corresponding H\&E images in (a) indicating that our classification results generalize beyond annotated tissue regions. The correspondence between RF and H\&E is poor as expected from the AUC values. }
    \label{fig:results}
\end{figure}

Figure \ref{fig:results} presents results when the RF and CNN classification models are used to segment  ovarian tissue cores, including regions outside annotated areas. O-PTIR data is consistent with H\&E stained microscopy data, whereas RF results show poor concordance. These results show that our CNN classification results extend beyond annotated data indicating effective tissue segmentation into epithelium and stroma. 

\begin{figure}[hbt]
    \centering
 \includegraphics[width=\linewidth]{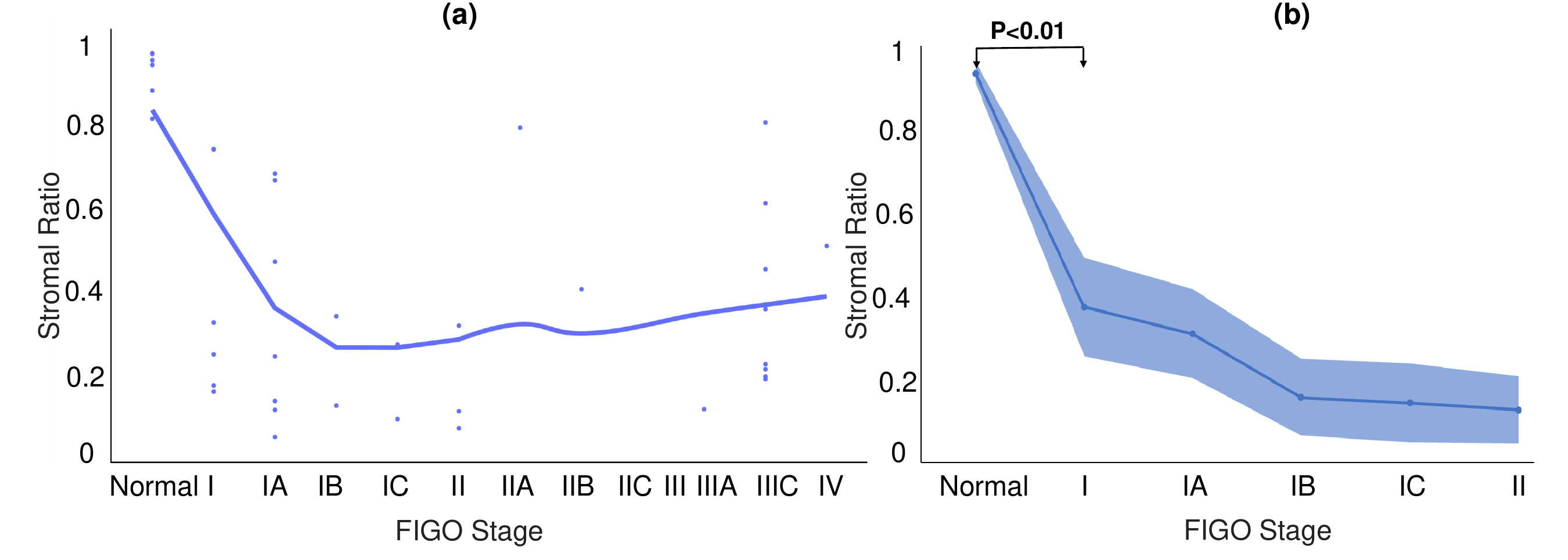}
  \includegraphics[width=\linewidth]{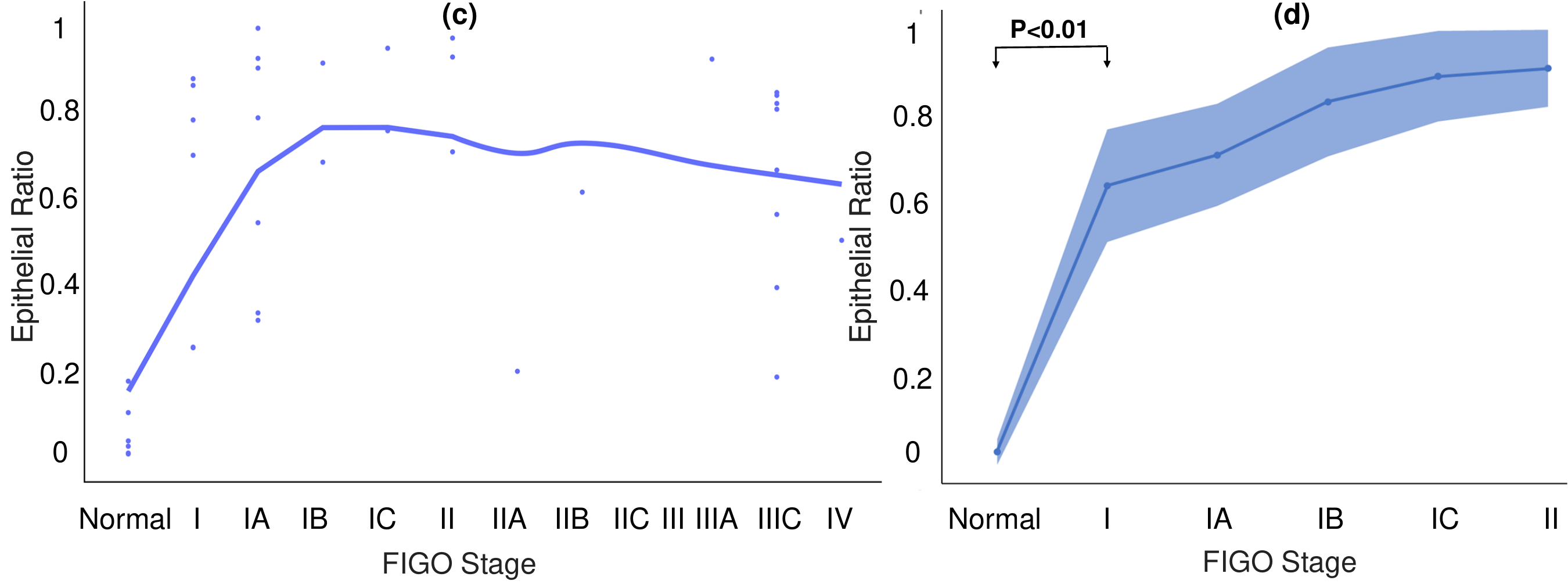}
    \caption{Stromal ratio (SR) and Epithelial ratio (ER) are plotted as a function of the pathologist-assigned FIGO stage of ovarian cancer. These ratios are calculated by dividing the number of pixels classified as stroma or epithelium by the total number of pixels in each biopsy core. The trend line in plots (a) and (c) show a nonlinear fit for SR and ER as a function of the cancer stage. These trend lines display a substantial reduction in SR and an increase in ER from normal to grade II, but do not change appreciably from grades II to IV. There is a statistically significant ($P<0.01$) decrease in SR and a significant ($P<0.01$) increase in ER during the early stage of cancer from normal tissue until stage II as presented in (b) and (d). The graphs show the mean SR and ER vs. early cancer grade. The error bands correspond to one standard deviation (SD).}
    \label{fig:ratioplot}
\end{figure}

We can utilize classification results to facilitate early ovarian cancer detection.  Figure \ref{fig:ratioplot} presents plots of stromal ratio (SR) and epithelial ratios (ER) as a function of Federation Internationale de Gynecolgie et d'Obstetrique (FIGO) cancer stage. SR and ER showed a nonlinear relationship with the FIGO stage, and the curve fitting function was calculated using nonlinear least-squares on the ratios. We observe a sharp reduction in SR with increasing cancer grade until grade II and a subsequent flattening of the curve. Figure \ref{fig:ratioplot}(b) emphasizes the SR trend in early cancer stages: normal, grade I, IA, IB, IC, and II. There is a statistically significant ($P<0.01$) reduction in SR from normal to grade I, illustrating its effectiveness as an early detection biomarker. A complimentary trend is observed in the ER metric in (c). A statistically significant ($P<0.01$) increase in ER is observed with progressively worsening early stage cancer in (d). Pathologists have {\it qualitatively} observed changes in the relative quantities of stroma and epithelium in early cancer stages. Our technique {\it quantifies} the number of epithelial and stromal pixels, thereby enabling measurement of {\it quantitative} trends in the aforementioned metrics in the early stages of ovarian cancer. 

\section{Discussion}
FTIR imaging measures tissue absorbance across all mid-IR wavenumbers and constructs a hyperspectral data cube. The technology does not allow the measurement of individual wavenumbers. O-PTIR uses a tunable QCL to measure tissue absorbance at discrete wavenumbers. We can measure only a subset of mid-IR wavenumbers relevant to a specific application, thereby reducing data collection time. On the other hand, O-PTIR uses a pump-probe architecture to obtain super-resolution images. This improved resolution results in an increase in the quantity of data and a corresponding increase in data collection time for each image. The effect of the improvement in resolution on tissue classification and the tradeoff between spatial and spectral resolution have not been studied until this paper. Our results demonstrate that we can maintain excellent tissue classification accuracy by reducing the number of bands and increasing the spatial resolution. This work presents a framework for making spatial-spectral tradeoffs in spectroscopic imaging while retaining good tissue segmentation accuracy.


Deep learning is used routinely in image classification.\cite{jiang2020emerging} However, it's application to hyperspectral data is limited.\cite{berisha2019deep} Furthermore, it has never been applied to super-resolution hyperspectral data, and our paper is the first to demonstrate efficacy. Hyperspectral data being three-dimensional (3D) requires a large memory bandwidth. Super-resolution images have finer spatial details and require larger convolutional kernels to identify the same area as FTIR, increasing computational costs and making classification more challenging. We have optimized our novel deep-learning architecture to achieve an excellent tissue segmentation AUC of 0.98 despite these challenges. 

Our results are obtained on 74 independent cancer patient cores and are statistically robust. Prior work often utilizes pixels within the same set of tissue cores during classification.\cite{kumar2013change, wang2021oral} This can lead to misleading results since the machine learning algorithm may learn features that correspond to specific patient traits that are challenging to generalize beyond the current dataset. We perform training and validation on mutually exclusive patient cores, achieving robust, generalizable results that enhance scientific rigor and reproducibility. 

Results in Figure \ref{fig:roc} show a significant improvement in efficacy between RF (AUC=0.58) and CNNs (AUC=0.98) on the same O-PTIR data. 
Since CNNs not only utilize spectroscopic information, but also extracts spatial data, these results highlight the advantages of combining spatial and spectral features. This work builds on prior spatial-spectral FTIR classification work\cite{pounder_development_2016} and affirms the validity of this research approach. 

The increased spatial resolution of O-PTIR leads to larger within-class spectral variation. Spectra in FTIR imaging are averaged over $\approx 5 \mu m$ pixel, which is approximately the size of one cell. On the other hand, spectra in O-PTIR correspond to more localized ($0.5 \mu m$) sub-cellular features such as cell nucleus or Golgi apparatus, which have disparate biochemical constituents. This leads to a larger spectra variation in O-PTIR even within the same tissue class. A large within-class variance can be a potential disadvantage in tissue segmentation and analysis. However, our data analysis approach that combines spatial-spectral features turns this variation into an advantage. 


The performance of machine learning algorithms depends critically on the quality and quantity of annotated data. Since annotations are performed using images of stained adjacent-sections that are several microns away from the MIRSI section, this imposes limitations on labeling accuracy. We mitigate annotation errors by limiting our labeling to unambiguous tissue areas and avoiding class boundaries. Furthermore, the alignment of images from adjacent sections is challenging,\cite{mankar2021multi} and we obtained the best fit through manual adjustment. The five wavenumbers that we chose for O-PTIR imaging offer good classification performance, but optimizing the set of wavenumbers could lead to improved performance. We will explore this optimization and the effects of improved spatial resolution on identifying other tissue subtypes in future work involving multi-class segmentation.

To our knowledge, this is the first large scale analysis of ovarian cancer tissue using mid-IR spectroscopic imaging. This analysis affords quantitative insights into ovarian cancer. Pathologists utilize the extent of epithelial infiltration into stroma and the relative proportion of stroma or epithelium to the rest of the tissue to subjectively assess cancer grade. Since our approach can precisely quantify the number of pixels of these subtypes, we can {\it quantify} these assessments and observe trends in a reliable manner. Furthermore, we analyze 74 cancer patients, enabling statistically robust analysis. Figure \ref{fig:ratioplot} (a) and (b) show that there is a statistically significant ($P<0.01$) reduction in the stromal ratio (SR) between normal tissue (SR $\approx 0.9$) and early stage (grade I - SR $< 0.4$) ovarian cancer. SR  Furthermore, SR reduces from grade I to II and then shows no appreciable change from grade II to IV. The grades are obtained directly from Biomax. A complimentary trend is observed in the epithelium ratio (ER) in (c) and (d). ER for normal tissue is $\approx 0.1$ and that for grade I cancer is $>0.6$. These results illustrate the utility of quantitative tissue classification in cancer diagnosis. SR and ER are quantitative   biomarkers for early stage ovarian cancer diagnosis and will be explored in greater detail in future work. 


\section{Conclusion}
MIRSI is an emerging technology that has the ability to revolutionize digital histopathology. Significant progress has been made in overcoming the technological challenges impeding its clinical adoption. O-PTIR solves the spatial resolution challenge of prior FTIR imaging technology, enabling label-free sub-cellular tissue investigation. In this work, we present the first label-free, automated histological classification of ovarian tissue subtypes using MIRSI. We show that the improved spatial resolution allows us to make fewer spectral band measurements and still achieve reliable tissue segmentation with an AUC of 0.98. These results are enabled using a novel deep-learning architecture optimized for MIRSI data. The results are statistically robust with validation over 74 cancer patients and 60 million data points. We utilize tissue classification and propose new quantitative biomarkers for early ovarian cancer diagnosis. The combination of deep learning and quantitative biochemical measurements using MIRSI enables numerically precise evaluation of previously subjective assessments, improving the rigor and reproducibility of histopathology. O-PTIR also performs measurements in back-reflection geometry, making the instrument easy to use on diverse tissue samples and facilitates future clinical translation. 




\medskip


\section*{Author Contributions}

\textbf{CCG}: Data curation, Visualization, Formal analysis, Investigation, Methodology, Software. \textbf{MB}: Data curation, Visualization, Investigation, Writing- Original draft preparation. \textbf{RM}: Methodology, Investigation. \textbf{NK}: Resources,  Data curation. \textbf{SC}: Data curation, Formal analysis, Validation. \textbf{YG}: Data curation, \textbf{JL}: Resources, Data Curation. \textbf{AKS}: Conceptualization, Supervision. \textbf{DM}: Supervision, Writing- Reviewing and Editing. \textbf{SB}: Investigation, Methodology, Software, Validation. \textbf{RR}: Conceptualization, Funding acquisition, Project administration, Resources, Supervision, Writing- Reviewing and Editing.

\section*{Conflicts of interest}
There are no conflicts to declare.

\section*{Acknowledgements}
This work is supported in part by the NLM Training Program in Biomedical Informatics and Data Science T15LM007093 (RM, RR), the Cancer Prevention and Research Institute of Texas (CPRIT) \#RR170075 (RR), National Institutes of Health \#R01HL146745 (DM), the National Science Foundation CAREER Award \#1943455, NLM Training Program in Biomedical Informatics and Data Science \#T15LM007093 SB), the NLM 2019 Data Science and Biomedical Informatics Undergraduate Summer Research Program (MB) grant 3T15LM007093-27S1,  National Institutes of Health P50 CA217685 (AKS), P30CA016672 (AKS), American Cancer Society Research Professor Award (AKS), Frank McGraw Memorial Chair in Cancer Research (AKS), Moon Shot in Ovarian Cancer (AKS).



\balance


\bibliography{refs} 
\bibliographystyle{rsc} 

\end{document}